\documentclass[a4paper,noindent]{JHEP3}
\usepackage{amsmath}
\usepackage{multirow}
\usepackage{mathenv}
\usepackage{array}
\usepackage{cite}
\usepackage{color}
\let\normalcolor\relax

\newcommand{\bea}{\begin{eqnarray}}
\newcommand{\eea}{\end{eqnarray}}
\newcommand{\bean}{\begin{eqnarray*}}
\newcommand{\eean}{\end{eqnarray*}}

\def\ket#1{\left| #1\right\rangle}

\def\gb #1{ \left\langle #1 \right]}

\def\cb #1{ \left[ #1 \right]}

\def\vev#1{\left\langle #1 \right\rangle}

\def\Label#1{\label{#1}}

\newcommand{\be}{\begin{equation}}
\newcommand{\ee}{\end{equation}}
\newcommand{\bd}{\begin{displaymath}}
\newcommand{\ed}{\end{displaymath}}

\newcommand{\Spapb}[3]{\langle  {#1} | {#2} | {#3} ]}

\title{Single Cut Integration}
\author{Ruth Britto, Edoardo Mirabella\\
Institut de Physique Th\'eorique, CEA-Saclay, F-91191, Gif-sur-Yvette
cedex, France \\
Email: \email{ruth.britto@cea.fr,  edoardo.mirabella@cea.fr}}

\abstract{
We present an analytic technique for evaluating single cuts for one-loop integrands, where exactly one propagator is taken to be on shell. 
 Our method  extends 
 the double-cut integration formalism of one-loop amplitudes to the single-cut case.   We argue that single cuts give meaningful information about amplitudes when taken at the integrand level.
We discuss applications to the computation of tadpole coefficients.
}

\preprint{IPhT-t10/163}

\begin{document}

\section{Introduction}

Precision calculations in gauge theories such as QCD are needed for observations in hadron collider experiments and are an important  motivation for studies of higher-order scattering in general.    Recently there have been rapid developments in computational techniques, many centered around so-called unitarity methods \cite{Bern:1994zx,Bern:1994cg}. 

 The idea underlying the unitarity methods is to constrain amplitudes by their branch cuts in various channels.   The constraints from cuts are sufficient only for certain classes of amplitudes, such as one-loop amplitudes in massless theories \cite{vanNeerven:1985xr}.  Unitarity cuts are evaluated by the Cutkosky rules \cite{Cutkosky:1960sp}, which put two propagators on shell.  Rather than completing the dispersion integral at this point, unitarity methods store the imaginary part of the amplitude obtained this way, and accumulate similar information from all possible cuts in analytic continuations to other kinematic regions.

Putting more than two propagators of a one-loop amplitude on shell gives a ``generalized'' unitarity cut \cite{Bern:1997sc,Bern:2004ky,Britto:2004nc,Mastrolia:2006ki,Forde:2007mi,Ossola:2006us,Ellis:2007br,Kilgore:2007qr,Giele:2008ve,Berger:2008sj,Badger:2008cm,Ellis:2008ir}.  It isolates the part of the amplitude from diagrams containing the specific propagators being cut.  An advantage of such cuts is that the delta functions that are used to place the propagators on shell effectively reduce the dimension of the remaining integral.  So these generalized cuts are relatively easy to evaluate but give information about a correspondingly smaller part of the full amplitude.  However, generalized unitarity can also be taken in the other direction, by cutting just one propagator.  This is the ``single cut'' we discuss in this paper.  Our focus differs from previous studies \cite{NigelGlover:2008ur,CaronHuot:2010zt,Catani:2008xa,Bierenbaum:2010cy} in that we are interested in the explicit evaluation of the 4-dimensional single-cut integral, with standard Feynman propagators, with a view towards computing tadpole coefficients in amplitudes with internal masses.  

One-loop amplitudes have an expansion in terms of master integrals, which are scalar box, triangle, bubble and tadpole integrals, with coefficients that are rational functions of the kinematic invariants \cite{Brown:1952eu, Melrose:1965kb, Passarino:1978jh, Hooft1978xw, vanNeerven:1983vr, Stuart:1987tt,Stuart:1989de,vanOldenborgh:1989wn,Bern:1992em, Bern:1993kr,Fleischer:1999hq,Binoth:1999sp,Denner:2002ii,Duplancic:2003tv,Denner:2005nn,Ellis:2007qk}.
In practice, unitarity methods operate by pattern matching cuts of amplitudes with cuts of master integrals.  Tadpole integrals lack cuts in physical channels (double cuts), but we hope that single cuts should be useful to calculate their coefficients.  In principle, single cuts should give information about all parts of the amplitude.  We study the single cuts of master integrals and find that the tadpole gives a rational value, while the others have purely logarithmic single cuts.  Therefore the tadpole coefficients can be targeted by discarding all logarithms.  
Other approaches to analytic computation of tadpole coefficients have been proposed, using universal UV and IR divergent behavior \cite{Bern:1995db,Badger:2008za} or introducing an auxiliary propagator \cite{Britto:2009wz}.

Inspired by the formalism for explicit evaluation of double cuts \cite{Britto:2005ha,Britto:2006sj,Anastasiou:2006jv,Britto:2006fc,Anastasiou:2006gt,Britto:2007tt,Britto:2008vq,Feng:2008ju,Mastrolia:2009dr}, where the two-dimensional (or ($D-2$)-dimensional) integral is performed algebraically by the Cauchy residue theorem, we rewrite our loop momentum in terms of spinor variables, which can in turn be exchanged for a complex variable and a real parameter.  Since we now have fewer cut constraints than in a double cut, we are not able to evaluate the integral over the real parameter trivially.  In fact, we leave it unevaluated.  We find that it is sufficient to work at the integrand level; moreover, the full single cut integral will typically diverge.  (Supersymmetric theories are a notable exception: there, single cuts are well defined \cite{CaronHuot:2010zt} and can be used to check expressions for planar multi-loop amplitudes \cite{ArkaniHamed:2010kv}.)  The integral over the complex variable is addressed by the Generalized Cauchy Theorem, invoked similarly in \cite{Mastrolia:2009dr}.  If we are computing single cuts to find tadpole coefficients, we see another major difference compared to double cuts.  While double cut evaluations were a matter of evaluating residues at poles, here it is the contour integral part of the formula that dominates the tadpole contribution, so we do not compute residues at all.

 Working at the integrand level is important because functions of the loop momentum can have non-vanishing single cuts even if they integrate to zero \cite{CaronHuot:2010zt}.  Therefore, a proper expansion of the integrand from a Feynman diagram includes all such terms, which have been thoroughly classified in the context of four-dimensional reduction \cite{Ossola:2006us,Ossola:2007bb}.

The paper is organized as follows. In Section 2, we explain the phase space integration for single cuts and evaluate the cuts for master integrals.  In Section 3, we present the single cuts of the first few types of integrands with tensor numerators.  In Section 4, we extract the tadpole coefficients in some sample tensor integrands.  In Section 5, we discuss the modifications necessary in the case of massless external legs (which are vanishing Gram determinants for bubble integrals).  
The appendix contains further details on the phase space parametrization.

\section{Evaluation of the single cut}
\label{Sec:SingleCut}
Our starting point is the one-loop integrand, 
\be
I = \frac{N(k)}{D_0 D_1 \cdots D_k},
\ee
where $N(k)$ is a polynomial in the loop momentum $k$, and the denominator factors are
\be
D_i = (k-K_i)^2 - m_i^2.
\ee
The single cut is a singularity of the amplitude selecting single propagators. 
We define the 4-dimensional single-cut operator for a particular propagator $D_i$ 
to act on the integrand as
\bea
\Delta_{D_i}[I] \equiv 
\int d^4k~  \delta^{(+)}(D_i) \left( \frac{N(k)}{D_0 \cdots D_{i-1}D_{i+1} \cdots D_k}  \right).
\Label{operator-def}
\eea
The single cut must be applied to the  {\it integrand}, because there are non-vanishing contributions from so-called spurious terms (terms that vanish upon integration).  Working with the  integrand allows us to identify the particular propagator being cut.

To evaluate the single cut analytically, we introduce convenient variables and reference vectors following \cite{Cachazo:2004kj,Anastasiou:2006jv,Mastrolia:2009dr}.  First, we would like to exchange the original loop momentum variable $k$ for a null vector $\ell_1$ in order to make use of the spinor formalism \cite{Berends:1981rb,De Causmaecker:1981bg,Kleiss:1985yh,Xu:1986xb,Gunion:1985vca}.
With respect to an arbitrary Lorentz vector $K$ satisfying $K^2 \neq 0$ and $K_0 >0$, we decompose the loop momentum variable,
\bea
k = \ell_1 + \xi K,
\Label{ellone}
\eea
where $\xi$ is a scalar factor taking a value such that $\ell_1$ is null.  This condition is implemented by another delta function and an integral over $\xi$,
\bean
\int d^4k~ (\bullet) = \int d\xi~ \int d^4\ell_1~ \delta^{(+)}(\ell_1^2)(2\ell_1 \cdot K)(\bullet).
\eean
 Let us now look at  the single cut integral, (\ref{operator-def}).  For simplicity, suppose we have redefined the loop momentum so that $K_i=0$.  We comment further on allowable redefinitions below.  Also, let $m=m_i$.
 Applying the  change of variables (\ref{ellone}), we have
\bea
 \int d^4k~ \delta^{(+)}(k^2 - m^2) ( \bullet ) 
&=& \int d\xi~ \int d^4\ell_1~ \delta^{(+)}(\ell_1^2)
\frac{ 2\ell_1 \cdot K }
{ \sqrt{\Delta}} 
\delta(\xi-\bar\xi)
(\bullet),
\Label{1st-ch}
\eea
where
\bean
\Delta=(2\ell_1\cdot K)^2 + 4K^2 m^2,
\qquad
\bar\xi=\frac{-2\ell_1\cdot K + \sqrt{\Delta}}{2K^2}.
\eean
We continue by exchanging the integration of $\ell_1$ over the lightcone for an integral over the complex plane and a real parameter $t$.  
First, we express the vector $K$ as a sum of two null vectors, $p$ and $q$, which will later allow us to integrate over familiar complex variables instead of spinors.  Since our choice of $K$ was arbitrary, we can consider the choice of null $p$ and $q$ to be the starting point, making sure that $p \cdot q \neq 0$ and $p_0 + q_0 > 0$.
\bean
K=p+q, \qquad p^2=q^2=0.
\eean  
The replacement for the loop momentum is
\bea
\ell_1^\mu = t \left( p^\mu + z \bar{z} q^\mu + \frac{z}{2}\gb{q|\gamma^\mu|p}
- \frac{\bar z}{2}\gb{p|\gamma^\mu|q} \right) .
\eea
Since $2p\cdot q = K^2$ and $2 \ell_1 \cdot K = t(1+z\bar{z})K^2$, the integral measure becomes
\bea 
 \int d^4k~ \delta^{(+)}( k^2 - m^2) ( \bullet ) 
= \int d\xi 
\int_0^\infty  \frac{dt}{4} \int \left(i dz \wedge d\bar{z} \right)
\frac{K^2 t^2 (1+z\bar{z}) ~\delta(\xi-\bar\xi) }
{ \sqrt{ t^2 (1+z\bar{z})^2+u }} 
(\bullet), 
\Label{cut-var}
\eea
where now
\bea
\bar\xi &=& \frac{u}{2}
\frac{1}
{\sqrt{t^2 (1+z\bar{z})^2+u } +  t(1+z\bar{z}) },
\Label{xibar}
\eea
and we have defined 
\bean
u \equiv \frac{4 m^2}{K^2}.
\eean
We will find it convenient to work in the limit $u \to 0$, equivalent to  choosing our arbitrary $K$ such that $K^2 \gg m^2$.  In fact, we will be able to set $u=0$ exactly in all the cuts we study.

Now, the $\xi$ integral can be performed immediately by the delta function substitution.  The complex integration over $z$ and $\bar{z}$ will be performed by the Generalized Cauchy Formula as described in \cite{Mastrolia:2009dr}.  That is, for the integrand $F(z,\bar{z})$ we construct a primitive $G(z,\bar{z})$ with respect to, say, $\bar{z}$. Let $D$ be a disk in the complex plane encompassing all poles of  $G(z,\bar{z})$ viewed as a function of $z$. Then
\bea
\int_D   F(z,\bar{z}) ~d\bar{z} \wedge dz
=\oint_{\partial D} dz~ G(z,\bar{z}) 
- 2 \pi i \sum_{{\rm poles}~ z_j } {\rm Res} \{G(z,\bar{z}),z_j \}. 
\Label{gcf}
\eea

In practice, we use $\Lambda$ to denote the radius of the disk $D$ and rewrite the complex variable in terms of polar coordinates:
\bean
z=re^{i\alpha}; \qquad D=\{(r,\alpha) ~|~ 0\leq r \leq \Lambda, ~0 \leq \alpha < 2 \pi \}.
\eean
Additional details on the phase space parametrization are given in the appendix.

The final integration over $t$ will not actually be carried out; we extract the information we need at the integrand level.  In fact, the integral over $t$ typically diverges.  It should be cut off both from above and below, and any further transformations should be consistent with such a cutoff.  For example, we allow linear shifts of the original loop momentum, but not global rescalings.  

Analytically, the $t$-dependence of the cut integrands turns out to be useful, as we can restrict our attention to leading or subleading terms in $t$ in order to compute tadpole coefficients.  

In our first change of variables for the single cut, (\ref{1st-ch}), we assumed that the cut propagator was already in the form $(k^2 - m^2)$.  In fact, it is good 
to redefine the loop momentum $k \to k + K_i$ in each term of the single cut operation (\ref{operator-def}) so that the delta function is $\delta^{(+)}(k^2 - m^2)$. Because the single cut integral diverges, this redefinition is most obviously valid in the limit of large $K^2$, which we will implement routinely.  Otherwise, we must pay attention to the details of the cutoffs.

We will now study the single cuts of the integrands of master integrals.

\subsection*{Single cut of tadpole}

The scalar tadpole is the simplest integrand allowing the single cut operation.  With the variables of (\ref{cut-var}),
\bean
\Delta_{D_0}\left[\frac{1}{k^2-m_0^2}\right]
&=& \int  \frac{dt}{4} \int \left(i dz \wedge d\bar{z} \right)
\frac{K^2  t^2 (1+z\bar{z}) }
{ \sqrt{  t^2 (1+z\bar{z})^2+u}}. 
\eean
One $\bar{z}$-primitive of the integrand is 
\bea
{\cal P}_1 \equiv
(1/z) K^2\sqrt{t^2 (1+z\bar{z})^2+u},
\Label{calI1}
\eea
 so the 
result of applying the Generalized Cauchy Formula (\ref{gcf})
is
\bean
(2\pi )K^2 
\left(\sqrt{t^2 (1+\Lambda^2)^2+u}
-\sqrt{t^2 +u} \right).
\eean
The factor of $i$ has dropped out in converting the differential form (see the appendix for details).

In the limit of vanishing $u$, the single cut of the tadpole is therefore
\bean
(2\pi )K^2 t \left( \Lambda^2 + \frac{u}{2t^2(1+\Lambda^2)}
-\frac{u}{2t^2} \right).
\eean
We omit writing the integral over $t$ (and the factor of 4 in the denominator), since it will not be necessary in this paper to carry out this integration.  In subsequent integrals we will also drop all subleading terms in $u$ so that the formulas are more manageable, while still sufficiently distinct.

\subsection*{Single cut of bubble}

Consider the integrand $1/(D_0D_1)$, and take the single cut of $D_0$.
\bean
\Delta_{D_0}\left[\frac{1}{(k^2-m_0^2)((k-K_1)^2-m_1^2)}\right]
&=& \int d^4k~ \delta^{(+)}(k^2 - m_0^2) 
\frac{1}{D_1}.
\eean
Now, under the integral with the delta function, 
\bea
D_1=f_1 - 2\ell_1 \cdot K_1 - 2 \bar\xi K \cdot K_1,
\Label{dk63}
\eea
where
\be
f_i \equiv K_i^2-m_i^2+m_0^2.
\label{eq:fi}
\ee
In the limit of vanishing $u$, we also have $\bar\xi \to 0$, so we 
can neglect the last term of (\ref{dk63}).\footnote{Indeed, one can check using (\ref{xibar}) that in the rest frame $K=(K_0,0,0,0)$, this term falls off as $1/K_0$.}
So we proceed with the replacement $D_1=f_1-t\gb{\lambda|K_1|\tilde\lambda}$, followed by expansion in the null vectors $p$ and $q$.  The single cut is
\bea
 \int \left(i dz \wedge d\bar{z} \right)
\frac{K^2  t  }
{{\cal F}(K_1,z,\bar{z})},
\Label{alger}
\eea
where we have defined
\bea
{\cal F}(K_i,z,\bar{z})
\equiv  f_i-t 
\left( \gb{p|K_i|p}+\bar{z}\gb{p|K_i|q}+z\gb{q|K_i|p}
+ z \bar{z} \gb{q|K_i|q} \right).
\Label{calF}
\eea
One $\bar{z}$-primitive of the integrand in (\ref{alger}) is given by
\bea
{\cal P}_2 \equiv
-\frac{K^2 \log {\cal F}(K_1,z,\bar{z})}
{\gb{p|K_1|q} + z \gb{q|K_1|q}}.
\Label{calI2}
\eea

\subsection*{Single cut of triangle and box}

The analysis of triangle and box integrands is similar to the bubble.
For the triangle integrand, $1/(D_0D_1D_2)$,  the single cut can be expressed in terms of the primitive, given by
\bea
{\cal P}_3 &=&
 \int d\bar{z}~
\frac{K^2  t  }
{{\cal F}(K_1,z,\bar{z}) {\cal F}(K_2,z,\bar{z})} \\
&=&
-\frac{K^2}{{\cal D}(K_1,K_2,z)}
\log \left(
\frac{{\cal F}(K_1,z,\bar{z})}{{\cal F}(K_2,z,\bar{z})}
\right).
\Label{calI3}
\eea
For the box $1/(D_0D_1D_2D_3)$, the $\bar{z}$-primitive is given by
\bea
{\cal P}_4 &=&
 \int d\bar{z}~
\frac{K^2  t  }
{{\cal F}(K_1,z,\bar{z}) {\cal F}(K_2,z,\bar{z}) 
{\cal F}(K_3,z,\bar{z})} \\
&=&
\frac{K^2 \gb{\lambda|K_1|q} \log{{\cal F}(K_1,z,\bar{z})} }
{{\cal D}(K_1,K_2,z) {\cal D}(K_3,K_1,z)}
+ \frac{K^2 \gb{\lambda|K_2|q} \log{{\cal F}(K_2,z,\bar{z})} }
{{\cal D}(K_2,K_3,z) {\cal D}(K_1,K_2,z)}
\nonumber \\ &&
+ \frac{K^2 \gb{\lambda|K_3|q} \log{{\cal F}(K_3,z,\bar{z})} }
{{\cal D}(K_3,K_1,z) {\cal D}(K_2,K_3,z)}.
\Label{calI4}
\eea
We have defined
\bean
{\cal D}(K_i,K_j,z) \equiv
f_j \gb{\lambda|K_i|q} - f_i \gb{\lambda|K_j|q}
-t \cb{pq}\vev{\lambda|K_iK_j|\lambda},
\quad {\rm where~~} \ket{\lambda}=\ket{p}+z\ket{q}.
\eean

\subsection*{Overview: master integrands and strategy for single cuts}
We have found the primitives ${\cal P}_1, {\cal P}_2, {\cal P}_3, {\cal P}_4$, which are associated to the various master integrands.  The single-cut calculation would be completed by the Generalized Cauchy formula,  (\ref{gcf}).  It turns out that in the limit $\Lambda \to \infty$, we can ignore the residues and restrict our attention to the closed line integral term.  In polar coordinates,
\bea
\oint_{\partial D } {\cal P}_n(z,{\bar z}) ~dz = 
i \int_0^{2\pi} d\alpha~ \Lambda e^{i\alpha} 
{\cal P}_n(\Lambda e^{i\alpha},\Lambda e^{-i\alpha}).
\eea
In the limit $\Lambda \to \infty$, the
 leading behavior of the integrands is
\bea
\Lambda e^{i\alpha}  {\cal P}_1 & \simeq &  \Lambda^2 t, \\
\Lambda e^{i\alpha}  {\cal P}_n  & \simeq &   \frac{\log( \Lambda^2)}{\Lambda^{n-2}}, ~~~~~~~ n=2,3,4. 
\eea
The higher-point integrands are suppressed by powers of $\Lambda$.  Moreover, among all these primitives, all rational terms come from tadpoles.  Tadpole primitives are purely rational, while the others are purely logarithmic.  Therefore, in an algorithm targeting tadpole coefficients, we will select  terms of the single cut with specific dependence on $\Lambda^2$. 

For fully generic integrands, we only need the single-cut operator selecting the terms proportional to $\Lambda^2 t$, which we denote by a bar.
\bea
 \bar \Delta_{D_i}\left [ I \right ]    &\equiv&  \Delta_{D_i}\left [ I \right ] \Big|_{\Lambda^2 t - \text{terms}}.
\label{Eq:DefOp}
\eea
For integrands with null external momenta, it is convenient to define some further refinements of the operator, as follows.
\begin{eqnarray}
\tilde\Delta_{D_0}\left [ I \right ]   &\equiv& \Delta_{D_0}\left [ I \right ] \Big|_{\Lambda^4 t^2 - \text{terms}}.  \nonumber \\
\hat  \Delta_{D_0}\left [ I \right ]   &\equiv& \Delta_{D_0}\left [ I \right ]\Big|_{ \log(\Spapb{q}{K_1}{q}\Lambda^2) - \text{terms}} .
\label{Eq:DefOp1}
\end{eqnarray}
In the case where the masses of the propagators are not all distinct, we may also want to collect the single cuts of all propagators with a given mass, as indicated by the subscript $m^2$.
\bea
\bar  \Delta_{m^2}\left [ I \right ]   &\equiv&  \sum_{i:m_i=m} \bar  \Delta_{D_i}\left [ I \right ] .
\label{Eq:DefOp2}
\eea
It is now clear how to distinguish single cuts of the various master integrands.  Since spurious terms give nonvanishing single cuts as well, we need to know them in more detail.  We now compute the single cuts of some integrands in general form.  The results will allow us to compute single cuts of spurious terms on one hand and the single cut of a total integrand expansion on the other.

\section{Single cut of general integrands}
\label{Sec:SingleGeneralIntegrands}
In listing the single-cut results of general  integrands, it is as convenient and more useful to list them in terms of general numerators and denominators. A general integrand will take the form
\be
I_{n,p}(A_1, \ldots,  A_p; D_0, \ldots,  D_{n-1}) \equiv 
\frac{\prod_{i=1}^p  (2\, k\cdot A_i)}{\prod_{j=0}^{n-1} D_j },
\ee
since any appearance of the contraction $k^2$ in the numerator can be replaced by $m_0^2$ in the single cut.
Using  (\ref{cut-var}) in the limit $u \to 0$, we see that the single cut of $D_0$ gives
\bea
\Delta_{D_0} \left[ I_{n,p} \right] &=&  K^2  t^{p+1} \int_{0}^{2\pi} d\alpha \;
 G_{n,p}(A_1, \ldots,  A_p; D_0, \ldots,  D_n; \alpha), 
\Label{G-pri}
\eea
where we have defined 
\be
G_{n,p}(A_1, \ldots,  A_p; D_0, \ldots,  D_{n-1}; \alpha)  \equiv \Lambda e^{i \alpha}
\left [ \int d \bar z \frac{\prod_{i=1}^p  \langle \lambda  |A_i | \tilde \lambda ] }{\prod_{j=1}^{n-1}  
( f_j - t \langle  \lambda |K_j|\tilde \lambda ]  ) } \right ]_{z\to \Lambda e^{i \alpha}, \bar z\to \Lambda e^{-i \alpha} }.
\ee
The spinors $\langle \lambda |$ and $| \tilde \lambda]$ depend on $z$ and $\bar z$ as follows,
\be
\langle \lambda | =  \langle p  |  +   z  \langle  q | , ~~~~
| \tilde \lambda  ]  =  | p   ]  + \bar z  | q ]. 
\ee
 The single 
cut of $I_{n,p}$ is known once the integral  
$G_{n,p}$ has been computed.  We now  
compute the single cut of $I_{n,p}$ for several values of $(n,p)$.

\subsection*{Computation of $\mathbf{\bar\Delta_{D_0} \left[  I_{1,1} \right ]}$}
It is easy to show that
\be
G_{1,1}(A_1; D_0; \alpha)  =  \Lambda^2 \Spapb{p}{A_1}{p} +  \Lambda^3 e^{i \alpha} \Spapb{q}{A_1}{p} +  \frac{\Lambda^3}{2} e^{-i \alpha} \Spapb{p}{A_1}{q} + \frac{\Lambda^4}{2}  
 \Spapb{q}{A_1}{q} .
\ee
Therefore, paying attention to the powers of $t$ included in (\ref{G-pri}), we find the following values for the first two refinements of single-cut operators.
\begin{eqnarray}
\bar \Delta_{D_0} \left[  I_{1,1}(A_1;D_0) \right ] &=&  0 .\nonumber \\
\tilde\Delta_{D_0} \left[  I_{1,1}(A_1;D_0) \right ] &=&   \pi  K^2 \Lambda^4 (2 A_1 \cdot q)   t^2 .   %
 \Label{Eq:I11DT} 
\end{eqnarray}

\subsection*{Computation of $\mathbf{\bar\Delta_{D_0} \left[  I_{2,1} \right ]}$}
The $\Lambda^2t-$part of the single cut (\ref{G-pri}) is the $\Lambda^2/t$ part of $G_{2,1}$, which is
\bean
-\frac{\Lambda^2}{t} \frac{{A_1}\cdot{q} }{{K_1}\cdot{q} }. 
\eean
The single cut of $I_{2,1}$ is obtained after the trivial integration over $\alpha$, giving 
\be
\bar\Delta_{D_0} \left[  I_{2,1}(A_1;D_0, D_1) \right ] = 
- \frac{{A_1}\cdot{q} }{{K_1}\cdot{q} }   \; \Delta_{D_0} \left[  \frac{1}{D_0} \right ]. 
\label{eq:I21gen}
\ee
In the $u\to 0$ limit,  $\bar\Delta_{D_0} \left[  I_{2,1}(A_1;D_0, D_1) \right ]$ is therefore proportional to the single cut of the tadpole. 

For later convenience we compute  $\hat \Delta_{D_0}[I_{2,1}]$   and  $\bar\Delta_{m^2}[I_{2,1}]$ in the case where $K_1^2=0$ and $m_0^2 =m_1^2 =m^2$.
The logarithmic part of $G_{2,1}$ is a lengthy expression.  The logarithmic part of of the single cut of  $I_{2,1}$  is
\begin{eqnarray}
\hat\Delta_{D_0} \left[  I_{2,1}(A_1;D_0, D_1) \right ] &=&  2\pi  K^2   t  \Bigg [ 
\frac{  \langle
   p|K_1|q] \langle q|A_1|p]}{ \langle q|K_1|q]^2}+\frac{  \langle
   p|K_1|p] \langle q|A_1|q]}{ \langle q|K_1|q]^2}\nonumber \\
   &~& +    \frac{ \langle
   p|A_1|q] \langle q|K_1|p]}{ \langle q|K_1|q]^2}-\frac{2  \langle
   p|K_1|q] \langle q|A_1|q] \langle q|K_1|p]}{ \langle
   q|K_1|q]^3} 
 -\frac{  \langle p|A_1|p]}{ \langle q|K_1|q]} \Bigg ] 
\nonumber \\
&   \equiv&  \sum_{a}  c_a(t) v_a^{\mu} A_{1\mu}. 
   \Label{Eq:I21DH}
\end{eqnarray}
The last line is simply an abbreviation for the expression, which will be convenient shorthand in one of the examples we give in Section \ref{sec:deg}.
In that case where  $K_1^2=0$  and $m_0^2 =m_1^2 =m^2$, we also have
\be
\bar \Delta_{m^2} \left[  I_{2,1}(A_1;D_0, D_1) \right ]  =
\bar \Delta_{D_0} \left[  I_{2,1}(A_1;D_0, D_1) \right ]  + \bar \Delta_{D_1} \left[  I_{2,1}(A_1;D_0, D_1) \right ]  = 0.
\Label{Eq:I21DM}
\ee
The single cut $\bar \Delta_{D_1}$ is computed redefining the loop momentum $k \to k+K_1$, as described in Section~\ref{Sec:SingleCut}.

\subsection*{Computation of $\mathbf{\bar\Delta_{D_0} \left[  I_{2,2} \right ]}$}
The expression of $G_{2,2}$ is rather complicated and it is not shown here.   
We obtain 
$\bar\Delta_{D_0} \left[  I_{2,2} \right ]$   
(or $\tilde\Delta_{D_0} \left[  I_{2,2} \right ]$) by  expanding $K^2 t^3 G_{2,2}$
in the limit $\Lambda \to \infty$ and then integrating the  $\Lambda^2t-$terms 
(or $\Lambda^4t^2-$terms).  The result is 
\bea
\Label{eq:I22gen}
\bar\Delta_{D_0} \left[  I_{2,2}(A_1,A_2;D_0, D_1) \right ]   &=&  
 -f_1 \frac{(A_1 \cdot q) (A_2 \cdot q)}{(K_1 \cdot q)^2}  \Delta_{D_0} \left [ \frac{1}{D_0} \right ] , \\
\tilde\Delta_{D_0} \left[  I_{2,2}(A_1,A_2;D_0, D_1) \right ]  &=& 
 - 2 \pi  K^2  \Lambda^4 \frac{ (A_1 \cdot q) (A_2 \cdot q)}{(K_1 \cdot q) } t^2 . 
\Label{Eq:I22DT}
\eea
For later convenience we  compute $\bar\Delta_{m^2}I_{2,2}$ in the case
where  $K_1^2=0$ and $m_0^2 =m_1^2 =m^2$, 
\begin{eqnarray}
\bar \Delta_{m^2} \left[  I_{2,2}(A_1,A_2;D_0, D_1) \right ]  &=&
\bar \Delta_{D_0} \left[  I_{2,2}(A_1,A_2;D_0, D_1) \right ]  + \bar \Delta_{D_1} \left[  I_{2,2}(A_1,A_2;D_0, D_1) \right ]  \nonumber \\
&=& - \{ (2 K_1\cdot A_1) \bar \Delta_{D_0} \left[  I_{2,1}(A_2;D_0,D_1) \right ]   +  (A_1 \leftrightarrow A_2) \}.
\label{Eq:I22DM}
\end{eqnarray}

\subsection*{Computation of $\mathbf{\bar\Delta_{D_0} \left[  I_{3,1} \right ]}$}
The integrand $K^2 t^2 G_{3,1}(A_1;D_0,D_1,D_2;\alpha)$  does not contain $\Lambda^2t-$terms. Therefore   $\bar\Delta_{D_0} \left[  I_{3,1} \right ]$ vanishes.   
\be
\bar\Delta_{D_0} \left[  I_{3,1}(A_1;D_0, D_1,D_2) \right ] = 0.
\ee
This result confirms the absence of tadpole integrals in the reduction of $I_{3,1}$.

\subsection*{Computation of $\mathbf{\bar\Delta_{D_0} \left[  I_{3,2} \right ]}$}
The expression of $G_{3,2}(A_1,A_2;D_0,D_1,D_2;\alpha)$ is rather complicated
and is not shown.  As in the previous cases, $K^2 t^3 G_{3,2}$ has to be expanded 
in the $\Lambda \to \infty $ limit. We obtain $\bar\Delta_{D_0} \left[  I_{3,2} \right ]$ by
 integrating the terms of $\mathcal{O}(\Lambda^2t)$, finding
\be
\bar\Delta_{D_0} \left[  I_{3,2}(A_1,A_2;D_0, D_1,D_2) \right ] =  
\frac{({A_1}\cdot{q})({A_2}\cdot{q}) }{({K_1}\cdot{q})({K_2}\cdot{q}) }   \; \Delta_{D_0} \left[  \frac{1}{D_0} \right ].  
\label{eq:I32gen}
\ee

\subsection*{Computation of $\mathbf{\bar\Delta_{D_0} \left[  I_{3,3} \right ]}$}
 We obtain $\bar\Delta_{D_0} \left[  I_{3,3} \right ]$ by  taking the 
  $\Lambda \to \infty $ limit of  $K^2 t^4 G_{3,3}$ 
 and integrating the  $\Lambda^2t-$terms, to get
\be
\bar\Delta_{D_0} \left[  I_{3,2}(A_1,A_2,A_3;D_0, D_1,D_2) \right ] = \sum_{i=1}^2 \, f_i 
\frac{ ({A_1}\cdot{q})({A_2}\cdot{q})({A_3}\cdot{q})  }{({K_1}\cdot{q})({K_2}\cdot{q})({K_i}\cdot{q})  } 
  \; \Delta_{D_0} \left[  \frac{1}{D_0} \right ].  
  \label{eq:I33gen}
\ee

\section{Computation of tadpole coefficients}
\label{ComTadCoeff}
In this section we show how the single cut allows the computation of tadpole coefficients in several examples of small integrands.

We will compute $a(0)$, the coefficient of the tadpole integral 
$A_0(m_0^2)$, by  the single cut operator $\bar \Delta_{D_0}$ defined
in (\ref{Eq:DefOp}). 
We look at one-loop  integrands of the type
\begin{equation}
\mathcal{I}_{n,p} \equiv \frac{\prod_{i=1}^p  (2\, k\cdot R_i)}{\prod_{j=0}^{n-1} D_j }. 
\end{equation}
As described in the setup, when we cut the propagator $D_0$, we set $K_0 =0$.  Any appearance of the contraction $k^2$ in the numerator should then immediately be replaced by $m_0^2$, so that this form of the integrand is in fact general.

For now, we assume that the Gram determinant of $\mathcal{I}_{n,p}$ is nonvanishing and that  the masses 
are non-degenerate.  When this is not the case, further modifications are necessary, which we address in the following section. 

We will describe the derivation of the tadpole coefficients for the integrands with $(n,p)=\{(2,1),(2,2),(3,2),(3,3) \}$.  We have also verified the result for $(n,p)=(4,3)$, but as this calculation does not involve any notable new features, we do not present it here.  We have used 
\verb+FeynCalc+ \cite{Mertig:1990an} to check our results against those from Passarino-Veltman reduction.

The idea underlying our procedure is to expand the integrand in such a way that spurious terms are easily recognized.  The Ossola-Papadopoulos-Pittau  (OPP) decomposition~\cite{Ossola:2006us} is well suited for this purpose.  We take all the coefficients of both spurious and physical terms as unknowns.  We then drop the physical terms except for the tadpole.  Single cuts of all remaining terms are evaluated using the general results of the previous section.  The tensor $\prod_i R_i$ is expanded in a basis constructed  from fixed vectors; a convenient choice includes external momenta and orthogonal vectors, as in the OPP classification.  Thanks to this expansion, the single-cut equation becomes a system of separate equations, which are  the coefficients of independent tensors.   In the following examples, we illustrate the derivation of the system 
and the solution of the tadpole coefficient.

\subsection*{Tadpole coefficient of $\mathbf{\mathcal{I}_{2,1}}$}
The starting point is the OPP decomposition of $\mathcal{I}_{2,1}$  given by
\begin{equation}
\mathcal{I}_{2,1} = \frac{a(0)}{D_0} +  \tilde b_{11}(01) \, \frac{2 k \cdot  \ell_7}{ D_0D_1} +   \tilde b_{21}(01) \, \frac{2 k \cdot  \ell_8}{ D_0D_1} +
 \tilde b_{0}(01) \, \frac{2 k \cdot  n}{ D_0D_1} + \cdots.
 \label{eq:I21start}
\end{equation}
Terms whose single cut contains no  $\Lambda^2t$ contribution are included in ``$\cdots$'' . These terms will be systematically neglected throughout this section.
The momenta  $n$, $\ell_7$ and $\ell_8$ are defined~\cite{Ossola:2006us} to satisfy the conditions
\begin{displaymath}
K_1 \cdot n = K_1 \cdot \ell_7 =K_1 \cdot \ell_8 = 0,~~~~~ n^2 = \ell_7 \cdot \ell_8 = -K_1^2,~~~~~ \ell_7^2 = \ell_8^2 =0.
\end{displaymath}
 Applying the single cut operator $\bar \Delta_{D_0}$ and using 
(\ref{eq:I21gen}), we get
\begin{eqnarray}
0 &=&\Bigg [  - \left (a(0)+\alpha_1 \right ) K_1^\mu  +  \left  ( \tilde b_{11}(01)-\alpha_3 \right )  \ell^\mu_7 \nonumber  \\ 
&+&  \left ( \tilde b_{21}(01) -\alpha_4  \right )  \ell^\mu_8 + \left (  \tilde b_{0}(01) -\alpha_2 \right )  n^\mu \Bigg ] \; \frac{ q_\mu}{ K_1 \cdot  q } \; 
\Delta_{D_0}\left [\frac{1}{D_0} \right  ]. 
\label{eq:I21main}
\end{eqnarray}
Here $\alpha_{i=1,\cdots,4}$ are the coordinates of  $R_1$ in  the basis $\{ K_1, n, \ell_7, \ell_8 \}$. Explicitly, they read as follows:
\begin{equation}
\alpha_1 = \frac{R_1 \cdot K_1}{K_1^2}, ~~~\alpha_2 = - \frac{R_1 \cdot n}{K_1^2},~~~\alpha_3 =  - \frac{R_1 \cdot \ell_8}{K_1^2}, ~~~\alpha_4= - \frac{R_1 \cdot \ell_7}{K_1^2}.
\end{equation}

From the relation~(\ref{eq:I21main}) we see our first example of a system of equations leading to the tadpole coefficient.  Since $q$ can be chosen arbitrarily, the expression inside the square brackets vanishes.  This implies that each of the factors  multiplying the basis vectors vanishes separately, giving four equations.  

In fact, we only need the first of these equations to get the tadpole coefficient; here we do not need to solve for any of the spurious coefficients.  The result is
\begin{equation}
a(0)+\alpha_1 = 0 ~~\Longrightarrow~~ a(0) = - \frac{R_1 \cdot K_1}{K_1^2}.
\end{equation}
This value is in accordance with the one  obtained using the Passarino-Veltman decomposition.

\subsection*{Tadpole coefficient of $\mathbf{\mathcal{I}_{2,2}}$}
For $\mathcal{I}_{2,2}$, again we use the OPP expansion, keeping only the terms with a non-vanishing single cut of $D_0$.  These are the tadpole along with spurious terms,
\begin{eqnarray}
\mathcal{I}_{2,2} &=& \frac{a(0)}{D_0} +  \tilde b_{11}(01) \, \frac{2 k \cdot  \ell_7}{ D_0D_1} +   \tilde b_{21}(01) \, \frac{2 k \cdot  \ell_8}{ D_0D_1} +
 \tilde b_{0}(01) \, \frac{2 k \cdot  n}{ D_0D_1}  \nonumber \\
 &+&   \tilde b_{12}(01) \, \frac{(2 k \cdot  \ell_7)^2}{ D_0D_1} +   \tilde b_{22}(01) \, \frac{(2 k \cdot  \ell_8)^2}{ D_0D_1}  + 
 \tilde b_{01}(01) \, \frac{(2 k \cdot  \ell_7) (2 k \cdot  n)  }{ D_0D_1} \nonumber \\
 &+& \tilde b_{02}(01) \, \frac{(2 k \cdot  \ell_8) (2 k \cdot  n)  }{ D_0D_1}  +  \tilde b_{00}(01)   \left [ 
 \frac{(2k\cdot n)^2}{D_0D_1}  \right . \nonumber  \\
 &-& \left .  \frac{(2k\cdot K_1)^2 -4 k^2 K_1^2}{3D_0 D_1} \right ]  + \cdots  .
 \label{eq:I22start}
\end{eqnarray}
We compute the single cut of both sides of (\ref{eq:I22start}) using (\ref{eq:I21gen}) and (\ref{eq:I22gen}). The outcome is
\begin{eqnarray}
0 &=&\Bigg [  - \left  (\frac{a(0)}{f_1}+\alpha_1 + \frac{\tilde b_{00}(01)}{3} \right ) K_1^\mu K_1^\nu  +   
\left ( \tilde b_{00}(01)-\alpha_2 \right )  n^\mu n^\nu +
\left ( \tilde b_{12}(01)-\alpha_3  \right )  \ell_7^\mu \ell_7^\nu 
 \nonumber  \\ &+&
\left ( \tilde b_{22}(01)-\alpha_4 \right )  \ell_8^\mu \ell_8^\nu +
\left ( \frac{\tilde b_{0}(01)}{f_1} - 2 \alpha_5 \right )  K_1^\mu n^\nu +
\left ( \frac{\tilde b_{11}(01)}{f_1} - 2 \alpha_7 \right )  K_1^\mu \ell^\nu_7
\nonumber \\  &+&
\left ( \frac{\tilde b_{21}(01)}{f_1} - 2 \alpha_8 \right )  K_1^\mu \ell^\nu_8 +
\left ( \tilde b_{01}(01)-2 \alpha_{13} \right )  n^\mu \ell_7^\nu 
\nonumber \\  &+&
\left ( \tilde b_{02}(01)-2 \alpha_{14} \right )  n^\mu \ell_8^\nu \Bigg ]  \; 
 \Delta_{D_0}\left [\frac{1}{D_0} \right  ] \;   \frac{ q_\mu q_\nu}{( K_1 \cdot q )^2} f_1 . 
\label{eq:I22main}
\end{eqnarray}
Here $f_1$ is defined according to equation (\ref{eq:fi}), while 
$\alpha_{i=1,\ldots,16}$ are obtained from the following decomposition of  $R_1^\mu R_2^\nu$ in a basis of independent tensors: 
\begin{eqnarray}
R_1^\mu R_2^\nu &=& 
\alpha_1 K^\mu_1K^\nu_1 + 
\alpha_2 n^\mu n^\nu  +
\sum_{i=7}^8 \left [  \alpha_{i-4}  \ell_i^\mu \ell_i^\nu \right ] +
\alpha_5 (K^\mu_1n^\nu + K^\nu_1 n^\mu) +
\alpha_6 (K^\mu_1n^\nu - K^\nu_1 n^\mu)  \nonumber \pagebreak[1] \\ 
&+& 
\sum_{i=7}^8 \left [   \alpha_{i} (K^\mu_1\ell_i^\nu + K^\nu_1 \ell_i^\mu) +
\alpha_{i+2} (K^\mu_1\ell_i^\nu - K^\nu_1 \ell_i^\mu)  \right ] + 
\alpha_{11} (\ell^\mu_7\ell_8^\nu - \ell^\nu_7 \ell_8^\mu) +
\alpha_{12}g_{\mu \nu} \nonumber \pagebreak[1]  \\
&+&
\sum_{i=7}^8  \left [  \alpha_{i+6}  (n^\mu\ell_i^\nu + n^\nu \ell_i^\mu) +
 \alpha_{i+8} (n^\mu\ell_i^\nu - n^\nu \ell_i^\mu) \right ]  .
\end{eqnarray}
The condition~(\ref{eq:I22main}) is fullfilled for any lightlike  $q$ only if the second rank tensor inside the
square brackets vanishes. 
The first independent tensor includes the tadpole coefficient, but also the spurious coefficient $\tilde b_{00}(01)$.  So we need the second independent tensor as well, but no others.
In particular,
\be
\left \{ \begin{array}{ll}
\displaystyle  \frac{a(0)}{f_1}+\alpha_1 + \frac{\tilde b_{00}(01)}{3} = 0  \\ [0.3ex]
\displaystyle  \tilde b_{00}(01)-\alpha_2 =0
 \end{array} \right.
 ~~~\Longrightarrow~~~
 a(0) = -f_1 \left (   \alpha_1 + \frac{\alpha_2}{3}   \right ).
\ee
Using the explicit expression of $\alpha_1$ and $\alpha_2$,
\begin{eqnarray}
\alpha_1 &=& 
\frac{3(K_1\cdot R_1)(K_1\cdot R_2)}{2 (K_1^2)^2}
-\frac{(R_1\cdot R_2)}{2 K^2_1} 
- \frac{(n\cdot R_1)(n\cdot R_2)}{2 (K_1^2)^2},  \nonumber \\
\alpha_2 &=& 
- \frac{(K_1\cdot R_1)(K_1\cdot R_2)}{2(K_1^2)^2}
+\frac{( R_1\cdot R_2)}{2K^2_1} 
+ \frac{3 (n\cdot R_1)(n\cdot R_2)}{2  (K_1^2)^2}, \nonumber 
\end{eqnarray}
we get the value of the tadpole coefficient,
\be
a(0) = \frac{f_1}{3 (K_1^2)^2} \bigg (   K_1^2 \; (R_1\cdot R_2)  -4   (K_1\cdot R_1) (K_1\cdot R_2)\bigg ). 
\ee

\subsection*{Tadpole coefficient of $\mathbf{\mathcal{I}_{3,2}}$}
We now find it convenient to vary the OPP decomposition slightly.  Our single-cut decomposition of  $\mathcal{I}_{3,2}$ reads  as follows,
\begin{eqnarray}
\mathcal{I}_{3,2}   &=& \frac{a(0)}{D_0} +  \sum_{i=1}^{2}  \tilde c_{i2}(012) \frac{(2k\cdot \ell_{i+2})^2}{D_0D_1D_2}  
+ \sum_{j=1}^2  \sum_{i=1}^2  \tilde b_{i1}(0j) \frac{2k\cdot \ell_{i+2}}{D_0D_j} 
 \nonumber \\
&+& 
\sum_{i=1}^2 \tilde b_0(0i)\left ( \frac{2k\cdot K_{3-i}}{D_0D_i} -\frac{K_1\cdot K_2}{K^2_i}  \frac{2k\cdot K_i}{D_0D_i} \right ) 
  + \cdots.
  \label{eq:I32start}
\end{eqnarray}
As in OPP, 
$\ell_3$ and  $\ell_4$  are lightlike momenta such that 
\be
\ell_j^2 = K_i \cdot \ell_j =0, ~~~~ \forall \; i \in \{1,2 \}, \; j \in \{3,4 \}. 
\label{eq:LL34}
\ee
The evaluation of the single cut is  performed using equations (\ref{eq:I21gen}),~(\ref{eq:I22gen})
and~(\ref{eq:I32gen}).
The result can be written as follows, 
\begin{eqnarray}
0 &=& \Bigg [ 
\sum_{i=1}^2 \left ( \tilde c_{i2}(012) -\gamma^0_{i+2i+2}   \right ) \ell_{i+2}^\mu \ell_{i+2}^\nu 
 - \left ( \tilde b_{11}(01)+ 2 \gamma^+_{23}   \right ) K_{2}^\mu \ell_{3}^\nu  
- \left ( \tilde b_{11}(02) +2 \gamma^+_{13}   \right ) K_{1}^\mu \ell_{3}^\nu \nonumber \\ 
&-& \left ( \tilde b_{21}(01) +2 \gamma^+_{24}   \right ) K_{2}^\mu \ell_{4}^\nu  
- \left ( \tilde b_{21}(02) +2 \gamma^+_{14}   \right ) K_{1}^\mu \ell_{4}^\nu  
-  \left (\tilde b_0(02)  + \gamma^0_{11} \right ) K_1^\mu K_1^\nu \nonumber \\
&-&  \left (\tilde b_0(01)  + \gamma^0_{22} \right ) K_2^\mu K_2^\nu 
+ \left ( a(0) - 2 \gamma^+_{12}    + \tilde b(01) \frac{K_1\cdot K_2}{K_1^2}   
+  \tilde b(02) \frac{K_1\cdot K_2}{K_2^2}  \right ) K_1^\mu K_2^\nu
\Bigg ] \nonumber \\
&\times&  \Delta_{D_0}\left [\frac{1}{D_0} \right  ]  \;
 \frac{q_\mu q_\nu}{( K_1 \cdot q )( K_2 \cdot q )} . 
\label{eq:I32main}
\end{eqnarray}
The coefficients  $\gamma^{\cdots}_{ij}$ are obtained from the decomposition of $R_1^\mu R_2^\nu$ in a basis of independent tensors,
\begin{eqnarray}
R_1^\mu R_2^\nu &=&  \gamma_{00} g_{\mu \nu} +  \sum_{i=1}^2 \left [ 
\gamma^{0}_{ii} K_i^\mu K_i^\nu + 
\gamma^{0}_{i+2 i+2} \ell_{i+2}^\mu \ell_{i+2}^\nu \right ]  
+ \gamma^{+}_{12} (K_1^\mu K_2^\nu + K_1^\nu K_2^\mu) 
\nonumber \\
&+&   
\sum_{i=1}^2 \sum_{j=3}^4 \left [ \gamma^+_{ij}  (K_i^\mu \ell_j^\nu + K_i^\nu \ell_j^\mu )  +
\gamma^-_{ij}  (K_i^\mu \ell_j^\nu - K_i^\nu \ell_j^\mu )   
\right ] \nonumber \\ 
 &+&  \gamma^{-}_{34} (\ell_{3}^\mu \ell_{4}^\nu -\ell_{3}^\nu \ell_{4}^\mu) +
   \gamma^{-}_{12} (K_1^\mu K_2^\nu - K_1^\nu K_2^\mu).
\end{eqnarray}
Eq~(\ref{eq:I32main}) implies that the following conditions have to be fulfilled, 
\be
\left \{ \begin{array}{lll}
\displaystyle  \tilde b_0(01)  + \gamma^0_{22} = 0,  \\ 
\displaystyle  a(0) - 2 \gamma^+_{12}    + \tilde b_0(01) \frac{K_1\cdot K_2}{K_1^2}   
+  \tilde b_0(02) \frac{K_1\cdot K_2}{K_2^2} =0,  \\
\displaystyle  \tilde b_0(02)  + \gamma^0_{11}  =0.
 \end{array} \right.
 \label{eq:SysI32}
\ee
The system~(\ref{eq:SysI32}) together with  the explicit expressions of  $\gamma^0_{11}$, $\gamma^0_{22}$ and $\gamma^+_{12}$, 
gives the tadpole coefficient of $\mathcal{I}_{3,2}$, 
\be
a(0) =     2 \gamma^+_{12}    + \gamma^0_{22} \frac{K_1\cdot K_2}{K_1^2}   
+  \gamma^0_{11} \frac{K_1\cdot K_2}{K_2^2} \nonumber \\
= \frac{\sum_{i,j=1}^2 b_{ij} (K_i \cdot R_1)(K_j \cdot R_2)    }{
K_1^2 K_2^2 \left (  (K_1 \cdot K_2)^2 - K_1^2 K_2^2   \right ) 
}.
\label{eq:ResI32}
\ee
 The factors $b_{ij}$ appearing in (\ref{eq:ResI32}) are defined as follows,
 \begin{displaymath}
b_{12}=b_{21}= - K_1^2 K_2^2, ~~~~~~
b_{11}= K_2^2 (K_1\cdot K_2 ),~~~~~~
b_{22}= K_1^2 (K_1\cdot K_2 ).
\end{displaymath}

The decomposition~(\ref{eq:I32start}) relies on the particular  structure of the numerator
of $\mathcal{I}_{3,2}$.  Being more general, the OPP decomposition  does not take 
advantage  of the knowledge of the numerator of $\mathcal{I}_{3,2}$. As a consequence, 
new spurious terms enter. They are of the type 
\be
\frac{(2 k \cdot P)(2 k \cdot Q)}{D_0D_i},~~~~~~~\frac{2 k \cdot P}{D_0},
\label{eq:SpurOPP}
\ee
with $P,Q \in \{K_1,K_2,\ell_3, \ell_4 \}$ and  $i=1,2$. 
The coefficients of the terms~(\ref{eq:SpurOPP}) 
vanish. This can be    explicitly shown 
using the operator  $\tilde \Delta_{D_0}$ defined in~(\ref{Eq:DefOp1}), which 
selects the 
 $\Lambda^4 t^4$-enhanced terms of the single cut of
 $\mathcal{I}_{3,2}$.

\subsection*{Tadpole coefficient of $\mathbf{\mathcal{I}_{3,3}}$}
This is the last example of a tadpole coefficient that we will describe.  We have done the analogous calculation for ${\mathcal{I}_{4,3}}$, but it does not introduce any notable new features.

Here again, although we could have started with the full OPP expansion or other variations, we can simplify the calculation by the particular expansion of $\mathcal{I}_{3,3}$ in the following tensor integrands:
\begin{eqnarray}
\mathcal{I}_{3,3}   &=& \frac{a(0)}{D_0} 
+ \sum_{i=3}^{4}  \tilde c_{iii} \frac{(2 k \cdot \ell_{i})^3}{D_0D_1D_2} 
+ \sum_{i=3}^4  \sum_{j=1}^2  \tilde c_{iij} \frac{(2 k \cdot \ell_{i})(2 k \cdot \ell_{i})(2 k \cdot K_j)}{D_0D_1D_2}  \nonumber \pagebreak[1]  \\
&+& \sum_{i=3}^4 
\sum_{j=1}^2 \sum_{k=j}^2
 \tilde c_{ijk} \frac{(2 k \cdot \ell_i)(2 k \cdot K_j)(2 k \cdot K_k)}{D_0D_1D_2} \nonumber  \pagebreak[1]   \\
&-&  \sum_{i,j=1}^2  \tilde c_{iij}  \left (   f_j \left [  \frac{(2 k \cdot K_i)}{D_0D_{3-i}} \right ]_{\text{sp.}} + 
\left [ \frac{(2 k \cdot K_i)^2}{D_0D_{3-i}}  \right]_{\text{sp.}}  \delta_{ij} \right )
+\cdots. \pagebreak[1]  
\label{eq:I33start}
\end{eqnarray}
Here we are using the definition of $\ell_3$ and $\ell_4$ given in (\ref{eq:LL34}). The operator 
$[ \cdots ]_{\text{sp.}}$ selects  the spurious parts of its argument. The explicit 
expressions of the spurious part   can be read 
from equations~(\ref{eq:I21start}) and (\ref{eq:I22start}).
The single cut of both sides of equation (\ref{eq:I33start}) is computed
using   ~(\ref{eq:I21gen}),~(\ref{eq:I22gen}), and~(\ref{eq:I33gen}).  
 Using the decomposition of $R_{1}^\mu R_{2}^\nu R_{3}^\sigma$ in a basis of independent tensors,
\begin{eqnarray}
R_{1}^\mu R_{2}^\nu R_{3}^\sigma &=& 
\sum_{i,j,k=1}^2 \alpha_{ijk} K_{i}^\mu K_{j}^\nu K_{k}^\sigma+
\sum_{i=3}^4 \alpha_{iii} \ell_{i}^\mu \ell_{i}^\nu \ell_{i}^\sigma+
\sum_{i=3}^4 \Big ( 
\alpha^{+}_{i00} \ell_{i}^\mu g^{\nu \sigma}   +  \alpha^{+}_{00i} \ell_{i}^\sigma g^{\mu \nu} \nonumber \\
&+&     \alpha^{-}_{i00} \ell_{i}^\mu \left ( \ell_{3}^\nu \ell_{4}^\sigma - \ell_{4}^\nu \ell_{3}^\sigma\right ) \Big )  +
\sum_{i=1}^2 \Big (  \alpha^{+}_{i00} K_{i}^\mu g^{\nu \sigma} +
\alpha^{+}_{0i0} K_{i}^\nu g^{\mu \sigma}  +\alpha^{+}_{00i} K_{i}^\sigma g^{\mu \nu}   \Big )  \nonumber \\
&+& \sum_{i=1}^2 \Big (  \alpha^{-}_{i00} K_{i}^\mu  \left ( \ell_{3}^\nu \ell_{4}^\sigma - \ell_{4}^\nu \ell_{3}^\sigma\right )  +
\alpha^{-}_{0i0} K_{i}^\nu 
 \left ( \ell_{3}^\mu \ell_{4}^\sigma - \ell_{4}^\mu \ell_{3}^\sigma\right ) 
+\alpha^{-}_{00i} K_{i}^\sigma 
 \left ( \ell_{3}^\mu \ell_{4}^\nu - \ell_{4}^\mu \ell_{3}^\nu\right ) 
 \Big )  \nonumber \\
 &+&  \sum_{i,j=1}^2  \sum_{k=3}^4  \Big ( 
 \alpha_{ijk} K_{i}^\mu K_{j}^\nu \ell_{k}^\sigma +
 \alpha_{ikj} K_{i}^\mu K_{j}^\sigma \ell_{k}^\nu +
 \alpha_{kij} K_{i}^\nu K_{j}^\sigma \ell_{k}^\mu 
 \Big) \nonumber \\
 &+&  \sum_{i=1}^2  \sum_{j=3}^4  \Big ( 
 \alpha_{ijj} K_{i}^\mu \ell_{j}^\nu \ell_{j}^\sigma +
 \alpha_{jij} K_{i}^\nu \ell_{j}^\sigma \ell_{j}^\mu +
 \alpha_{jji} K_{i}^\sigma \ell_{j}^\nu \ell_{j}^\mu 
 \Big), 
 \end{eqnarray}
and equating the single cuts on both sides of equation (\ref{eq:I33start}), we get the  following relation:
\begin{eqnarray}
0 &=&
\sum_{m=1}^2  \Bigg \{ 
\Bigg [  
 \sum_{i=3}^{4}\hat  c_{iii}   \,   \ell^\mu_{i}   \ell^\nu_{i} \ell^\sigma_{i}  +
 \sum_{i=3}^{4}\sum_{j=1}^{2}\sum_{k=j}^{2} \hat c_{ijk} \,  \ell^\mu_i   K^\nu_j K^\sigma_k   
 + \sum_{i=3}^4 \sum_{k=1}^2 \hat c_{iik}  \ell^\mu_i\ell^\nu_i K^\sigma_k
\nonumber \\
&+& \sum_{i=1}^2  g_{ii;\; m}(\hat c)   \,   K^\mu_i   K^\nu_i K^\sigma_i   +
g_{3-m\;m;\; m}(\hat c)    \,   K^\mu_{3-m}   K^\nu_{3-m} K^\sigma_m  \nonumber\\
&+&  
 \frac{1}{2 f_m} K^\mu_m   K^\nu_m K^\sigma_{3-m}  \;    \Bigg ( a(0) - \sum_{i,j,k=1}^2 \alpha_{ijk} d_{ijk}  
 \nonumber  \\
&~&  +2 f_m \,  g_{m\; 3-m; \; m }(\hat c)  \Bigg ) \Bigg ]   \;
 \Delta_{D_0}\left [\frac{1}{D_0} \right  ]  \;
 \frac{ q_\mu q_\nu q_\sigma \; f_m}{( K_1 \cdot q )( K_2 \cdot q )( K_m \cdot q )}  \Bigg \}. 
\label{eq:I33main2}
\end{eqnarray}
The coefficients in the above relation have been defined for convenience of displaying the independent tensors.
We use the following abbreviations ($i\neq j \neq k \neq i$),
\begin{eqnarray}
\hat c_{iii}   &=&  \tilde c_{iii} -\alpha_{iii} \nonumber \\
\hat c_{iik}  &=&  \tilde   c_{iik}  -
(  \alpha_{kii} +  \alpha_{iki}  + 
\alpha_{iik}   ) \nonumber  \\
\hat c_{ijk}   &=& \tilde  c_{ijk}  - \left [
\alpha_{kji} + \alpha_{kij}  +\alpha_{ikj}
+(1-\delta_{jk}) (  \alpha_{jki} + \alpha_{jik}  +\alpha_{ijk}  )
\right ],
\end{eqnarray}
and the following definition of the totally symmetric coefficient $d_{ijk}$,
\begin{equation}
d_{iii}=  f_i \frac{ K_i \cdot K_{3-i}}{K^2_{3-i}}   - \frac{f_{3-i}}{3} \left (\frac{K_i^2}{K^2_{3-i}} - 4\frac{ (K_{3-i}\cdot K_i)^2}{(K^2_{3-i})^2} \right ),~~~~ 
d_{iij} = f_i + f_j \frac{K_i \cdot K_j}{K_j^2}. 
\end{equation}
The $g_{ij;\;m}(\hat c)$ are given by
\begin{eqnarray}
 g_{ij;\; m}(\hat c)&=& \delta_{ij} \hat c_{iii}    + (1- \delta_{ij}) \Bigg \{ 
\delta_{mj}  \hat c_{iij} -
 \delta_{mi} \Bigg [    \frac{K_1\cdot K_2}{K_j^2}   \hat c_{iii}   \nonumber \\
 &-&
 \frac{1}{3(K_i^2)^2} \left( K_1^2 K_2^2 -4 (K_1\cdot K_2)^2 \right )  \hat c_{jjj}  
  + \frac{K_1\cdot K_2}{K_i^2}   \hat c_{jji} 
 \Bigg ] \Bigg \}.
\label{Eq:Gmiij}
\end{eqnarray}
Since $q$ is arbitrary, we are led to the following relations.
\be
\left \{ \begin{array}{lllll}
\displaystyle  g_{11;\;2}(\hat c) = 0  \\ [0.3ex]
\displaystyle  g_{22;\;1}(\hat c) = 0  \\ [0.3ex]
\displaystyle f_1\, g_{11;\;1}(\hat c)+ f_2 \;  g_{12;\;2}(\hat c) = 0  \\ [0.3ex]
\displaystyle f_1\, g_{21;\;1}(\hat c)+ f_2 \;  g_{22;\;2}(\hat c) = 0  \\ [0.3ex]
\displaystyle  a(0) - \sum_{i,j,k=1}^2 \alpha_{ijk}  d_{ijk}  +  f_1  \, g_{12;\;1}(\hat c)  + f_2 \, g_{21;\;2}(\hat c)  =  0
 \end{array} \right.  ~,
\ee
which uniquely fix $a(0)$,
\be
a(0) = \sum_{i,j,k=1}^2 \alpha_{ijk}  d_{ijk}. 
\ee

We have observed that a different choice of the original integrand expansion, rather than (\ref{eq:I33start}), can lead to a larger linear system which is not completely solvable, yet there is still a unique solution for $a(0)$.  The underlying phenomenon is that different spurious terms can have the same single cut, so these terms should really be grouped together in the expansion.

\section{Massless external legs\label{sec:deg}}

As in other techniques such as 
Passarino-Veltman~\cite{Brown:1952eu, Melrose:1965kb, Passarino:1978jh, Hooft1978xw, vanNeerven:1983vr, Bern:1992em, Bern:1993kr,Denner:2005es, Denner:2005fg,Denner:2005nn} 
or 
Ossola-Papadopoulos-Pittau~\cite{Ossola:2006us, Ossola:2007bb}, we must modify our algorithm in the case of vanishing Gram determinants.  The most immediate case is the presence of a massless external leg in tensor bubble integrals. In this section we will focus on this class of integrals in the  case where the internal masses are the same, $m_0^2=m_1^2=m^2$. \medskip

New features appear in the computation of $a_{\mbox{\tiny tot}}$, the coefficient of $A_0(m^2)$.
First of all,  if $K_1$ is light-like,  it is not possible to complete a basis defining 
$n, \ell_7, \ell_8$. A suitable basis 
is given by $\{ K_1, K_2, \ell_7, \ell_8 \}$, which is composed of four light-like momenta such that
\be
K_1 \cdot  K_2  \neq  0 \neq   \ell_7 \cdot \ell_8, ~~~~~~~~
K_i \cdot \ell_j=0, ~~~~\forall \; i \in \{1,2\}, j \in \{7,8\}.   
\ee
Secondly,  the coefficients of $1 / D_0$ and $1/ D_1$ contribute to $a_{\mbox{\tiny tot}}$ since 
 \be
 \int d^4 k \frac{1}{D_0} = \int d^4 k \frac{1}{D_1} = A_0(m^2).
 \ee
Moreover the scalar  bubble and the tadpole are connected,
\be
B_0(0,m^2, m^2) =  \frac{A_0(m^2)}{m^2}-1,
\ee
so the bubble coefficient contributes to the 
total tadpole coefficient.  Tensor integrals
contracted  with $K_2$ are no longer spurious 
terms~\cite{Ossola:2007bb} and  they contribute to the 
total tadpole coefficient.  This can be easily understood   
looking at their explicit expression~\cite{Denner:2005nn},
\begin{eqnarray}
\int d^4 k \frac{(2 k\cdot K_2)}{D_0D_1} &=&
 B_1(0,m^2,m^2) (2 K_2 \cdot K_1) =  \frac{1}{2} \left ( 
\frac{A_0(m^2)}{m^2} - 1 \right )(2 K_2 \cdot K_1),
\nonumber  \\
\int d^4 k \frac{(2 k\cdot K_2)^2}{D_0D_1} &=&
B_{11}(0,m^2,m^2) (2 K_2 \cdot K_1)^2= \frac{1}{3} \left ( 
\frac{A_0(m^2)}{m^2} - 1 \right )(2 K_2 \cdot K_1)^2.
\end{eqnarray}
The tadpole coefficient $a_{\mbox{\tiny tot}}$ is obtained by summing the 
aforementioned contributions. These contributions  can be obtained by
applying the single cut operators defined in (\ref{Eq:DefOp}), (\ref{Eq:DefOp1}), and (\ref{Eq:DefOp2}).

\subsection*{Tadpole coefficient of $\mathbf{\mathcal{I}_{2,1}}$}
The  full integrand is expanded as 
\be
\mathcal I_{2,1} = \frac{a(0)}{D_0} +\frac{a(1)}{D_1} + \frac{b}{D_0D_1} +
\hat b_{0}(01) \frac{(2 k \cdot  K_2)}{D_0D_1}  +  \sum_{i=1}^2  \tilde b_{i1}(01) \frac{(2 k \cdot  \ell_{i+6})}{D_0D_1} + \cdots.
\Label{Eq:I21degINT}
\ee
The terms denoted  by ``$\cdots$'' are those such that 
$\bar \Delta_{m^2}[\cdots] = \bar \Delta_{D_0}[\cdots] = \hat \Delta_{D_0}[\cdots] =0$.
The tadpole coefficient is given by
\be
a_{\mbox{\tiny tot}} =  a(0) + a(1) + \frac{b}{m^2}  + \frac{\hat b_{0}(01)}{2m^2}(2 K_2 \cdot K_1). 
\Label{Eq:I21degM}
\ee
The sum $a(0) + a(1)$ can be obtained by applying $\bar \Delta_{m^2}$. Using equation (\ref{Eq:I21DM})
we see that only the tadpole terms survive, so 
\be
a(0) + a(1) = 0.
\Label{Eq:I21degA}
\ee
The coefficient $\hat b_0(01)$ is obtained by cutting the propagator $D_0$ and selecting the  $\Lambda^2 t$ terms. The 
outcome is  
\begin{eqnarray}
0 &=&  \Bigg [  -(a(0) + \alpha_1) K_1^\mu + (\hat b_0(01) - \alpha_2 )K_2^\mu \nonumber \\
&+& \sum_{i=1}^2 \left (  \tilde b_{i1}(01)
- \alpha_{i+2} \right ) \ell_{i+6}^\mu \Bigg ] \frac{ q^\mu}{{K_1}\cdot{q}} \Delta_{D_0} \left [ \frac{1}{D_0}\right ].
\Label{Eq:I21degR}
\end{eqnarray}
The parameters $\alpha_{1,\cdots,4}$ are the coordinates of $R_1$ in the basis $\{K_1,K_2,\ell_7,\ell_8 \}$, which 
read as follows:
\begin{equation}
\alpha_1 = \frac{R_1 \cdot K_2}{K_1\cdot K_2}, ~~~
\alpha_2 = \frac{R_1 \cdot K_1}{K_1\cdot K_2}, ~~~
\alpha_3 = \frac{R_1 \cdot \ell_8}{\ell_7 \cdot \ell_8}, ~~~
\alpha_4=  \frac{R_1 \cdot \ell_7}{\ell_7 \cdot \ell_8}.
\end{equation}
Since $q$ is arbitrary,  each coefficient of the momenta appearing in (\ref{Eq:I21degR}) has to vanish. In particular,
\be
\hat b_0(01) = \alpha_2 = \frac{R_1 \cdot K_1}{K_1\cdot K_2},
\Label{Eq:I21degB}
\ee
and $\tilde b_{i1}(01) = \alpha_{i+2} $.
Finally, the bubble coefficient $b$ can be obtained by using the operator $\hat \Delta_{D_0}$ defined 
in (\ref{Eq:DefOp1}). Using  equation (\ref{Eq:I21DH}), we get
\begin{eqnarray}
0 &=& \sum_a  \int \,  dt  \; c_a(t) v_{a\, \mu} \Bigg [ -\alpha_1 K_1^\mu + (\hat b_0(01) - \alpha_2 )K_2^\mu + \sum_{i=1}^2 \left (  \tilde b_{i1}(01)
- \alpha_{i+2} \right )  \ell_{i+6}^\mu   \Bigg ] 
\nonumber \\
&~& +  b \;  \Delta_{D_0} \left [ \frac{1}{D_0 D_1}\right ]  \nonumber \\
&=&  -\alpha_1 \sum_a  \int \,  dt  \; c_a(t)\; ( v_{a} \cdot K_1 )  + b\;  \Delta_{D_0} \left [ \frac{1}{D_0 D_1}\right ]  
\nonumber \\
&=& b\;  \Delta_{D_0} \left [ \frac{1}{D_0 D_1}\right ]   ~~\Longrightarrow~~ b =0.
\Label{Eq:I21degC}
\end{eqnarray}
The tadpole coefficient is obtained  from equation (\ref{Eq:I21degM}), using the results~(\ref{Eq:I21degA}),~(\ref{Eq:I21degB}), and~(\ref{Eq:I21degC}), and it is given by
\be
a_{\mbox{\tiny tot}} = \frac{(R_1\cdot K_1)}{m^2},
\ee
as expected from explicit reduction.

\subsection*{Tadpole coefficient of $\mathbf{\mathcal{I}_{2,2}}$}
Our final example is given to display the flexibility of single cut operations and the degrees of information available from cutting the same propagator.  We use several different refinements of the single cut, picking out terms with different dependence on $\Lambda$ and $t$, in order to collect the subset of the information required for the tadpole coefficient.

The integrand $\mathcal{I}_{2,2}$ is decomposed as follows,
\begin{eqnarray}
\mathcal{I}_{2,2} &=& \frac{a(0)}{D_0} + \frac{a(1)}{D_1} + \frac{b}{D_0D_1} + \sum_{i=1}^2 \left (  \tilde a_i(0) \frac{2k\cdot K_i}{D_0}   +
\tilde a_{i+2}(0) \frac{2 k \cdot \ell_{i+6}}{D_0}    \right ) \nonumber \\
&+&  
 \sum_{i=1}^2 \left (  \tilde b_{0i}(01) \frac{(2 k\cdot K_2)(2 k\cdot \ell_{i+6})}{D_0D_1}   +
\tilde b_{i2}(01) \frac{(2 k \cdot \ell_{i+6})^2}{D_0D_1}    \right ) \nonumber \\
&+& \hat b_{00}(01)  \frac{(2 k \cdot K_2)^2}{D_0D_1} + \cdots .
\end{eqnarray}
The terms denoted by  ``$\cdots$'' are not explicitly shown since   
$\bar \Delta_{m^2}[\cdots] = \tilde \Delta_{D_0}[\cdots] = \hat \Delta_{D_0}[\cdots] =0$.
The tadpole coefficient of $A_0(m^2)$ is given by
\be
a_{\mbox{\tiny tot}} =  a(0) + a(1) + \frac{b}{m^2}  + \frac{\hat b_{00}(01)}{3m^2}(2 K_2 \cdot K_1)^2. 
\Label{Eq:I22degM}
\ee
In the computation we will take advantage of the following decomposition of $R_1^\mu R_2^\nu$ in independent tensors,
\begin{eqnarray}
R_1^\mu R_2^\nu &=& 
\sum_{i=1}^2 \alpha_i K^\mu_iK^\nu_i + 
\sum_{i=7}^8 \left [  \alpha_{i-4}  \ell_i^\mu \ell_i^\nu \right ] +
\alpha_5 (K^\mu_1 K^\nu_2 + K^\nu_1 K^\mu_2) 
+ \alpha_6 (K^\mu_1K^\nu_2 - K^\nu_1 K^\mu_2)   \nonumber \\
&+& \sum_{i=7}^8 \left [   \alpha_{i} (K^\mu_1\ell_i^\nu + K^\nu_1 \ell_i^\mu) +
\alpha_{i+2} (K^\mu_1\ell_i^\nu - K^\nu_1 \ell_i^\mu)  \right ] + 
\alpha_{11} (\ell^\mu_7\ell_8^\nu - \ell^\nu_7 \ell_8^\mu) \nonumber \\
&+&
\alpha_{12}g_{\mu \nu} +
\sum_{i=7}^8  \left [  \alpha_{i+6}  (K^\mu_2\ell_i^\nu + K^\nu_2 \ell_i^\mu) +
 \alpha_{i+8} (K^\mu_2\ell_i^\nu - K^\nu_2 \ell_i^\mu) \right ] .
\label{eq:I22DMexpRR}
\end{eqnarray}
We apply the single cut and we select the $\Lambda^2 t$ terms using the operator $\bar\Delta_{m^2}$ and equation (\ref{Eq:I22DM}).
\begin{eqnarray}
0 &=& \Bigg [ 
\left(  \frac{a(0) + a(1) - 4 \alpha_{12}}{2K_1\cdot K_2}   - 2 \alpha_5   \right ) K_1^\mu+
2 (\hat b_{00}(01) - \alpha_2) K_2^\mu \nonumber \\
&+& \sum_{i=1}^2 (\tilde b_{0i}(01) - 2 \alpha_{i+12} )  \ell^\mu_{i+6} 
 \Bigg ] \frac{2 (K_1 \cdot K_2) q_\nu}{{K_1}\cdot{q}} \Delta_{D_0} \left [ \frac{1}{D_0}\right ].
 \Label{Eq:I22degSC}
\end{eqnarray}
Equation (\ref{Eq:I22degSC}) fixes the values of $a(0) + a(1)$ and of  $\hat b_{00}(01)$ to be
\begin{eqnarray}
\Label{Eq:I22degA}
a(0) + a(1) &=&    4 \alpha_{12} +  2 \alpha_5 (2K_1\cdot K_2),  \\
\hat b_{00}(01) &=& \alpha_2.
\Label{Eq:I22degB}
\end{eqnarray}
The value of $\tilde b_{0i}(01) $ is fixed as well, to be
\bea
\tilde b_{0i}(01) =2 \alpha_{i+12}.
\Label{fitie}
\eea
The value of $\tilde b_{i2}(01)$ is read off from the $\Lambda^4t^4$-enhanced terms of the 
single cut of $\mathcal{I}_{2,2}$. These terms are selected using the operator 
$\tilde \Delta_{D_0}$ and using  equations~(\ref{Eq:I11DT}) and~(\ref{Eq:I22DT}). The outcome can be written as follows,
\begin{eqnarray}
0 &=& \Bigg [  -(\tilde a_1(0) + \alpha_1 ) K_{1}^{\mu}K_{1}^{\mu}   - (\tilde a_2(0) +2 \alpha_5) K_{1}^{\mu}K_{2}^{\mu}  -
\sum_{i=3}^4 (\tilde a_{i}(0) + 2 \alpha_{i+4}) K_{1}^{\mu} \ell^\nu_{i+4} 
\nonumber \\
&~& + 
 \sum_{i=1}^2 (\tilde b_{0i}(01) - 2 \alpha_{i+12}) K_{2}^{\mu} \ell^\nu_{i+6} +
  \sum_{i=1}^2 (\tilde b_{i2}(01) - \alpha_{i+2} ) \ell_{i+6}^{\mu} \ell^\nu_{i+6} 
 \nonumber \\
&~& + (\hat b_{00}(01) - \alpha_2) K_{2}^{\mu} K_{2}^{\mu}    \Bigg ]  \frac{2 q_{\mu} q_\nu}{{K_1}\cdot{q}}  \pi K^2\Lambda^4  t^2~~
\nonumber \\
&\Longrightarrow& ~~ \tilde b_{i2}(01) = \alpha_{i+2}. 
 \Label{Eq:I22degS4}
\end{eqnarray}
 The logarithmically enhanced part of the single cut allows the computation of the bubble coefficient $b$. This is achieved by applying the $\hat \Delta_{D_0}$ operator defined in (\ref{Eq:DefOp1}).  Using equations~(\ref{fitie}) and (\ref{Eq:I22degS4}), we find
 \begin{eqnarray}
0 
&=& b\;  \Delta_{D_0} \left [\frac{1}{D_0D_1} \right ] - 
4 \alpha_{12} m^2  \Delta_{D_0} \left [ \frac{1}{D_0D_1} \right ]  
~~~\Longrightarrow~~~ b = 4 m^2 \alpha_{12}. 
 \Label{Eq:I22degC}
\end{eqnarray}
Here we have used the property
$
\hat \Delta_{D_0} I_{2,2}(K_1,A_2; D_0,D_1)   = 0$,
which holds in the case $f_1=0$.
  The tadpole coefficient in 
Equation (\ref{Eq:I22degM}) is then obtained using equations~(\ref{Eq:I22degA}),~(\ref{Eq:I22degB}) and~(\ref{Eq:I22degC}), 
\be
a_{\mbox{\tiny tot}} = 2\left ( 4  \alpha_{12} +\alpha_5  (2K_1\cdot K_2) \right )   + \alpha_2 \frac{(2 K_2 \cdot K_1)^2}{3m^2} =
2 R_1\cdot R_2 + \frac{(2R_1 \cdot K_1)(2R_2 \cdot K_1)}{3m^2}.
\Label{Eq:I22degAT}
\ee
In obtaining the result~(\ref{Eq:I22degAT}), we used some explicit values of the coefficients in the expansion (\ref{eq:I22DMexpRR}),
\be
 4  \alpha_{12} +\alpha_5  (2K_1\cdot K_2) = (R_1\cdot R_2),~~~~~~~
\alpha_2 = \frac{(R_1 \cdot K_1)(R_2 \cdot K_1)}{(K_1\cdot K_2)^2}.
\ee

\section{Conclusions and discussion}
We have seen how single cut integrals can distinguish scalar boxes, triangles, bubbles and especially tadpoles.    We have outlined a strategy to find the tadpole coefficients applying single cuts.  In particular
we have used the fact that the null vector $q$ was chosen arbitrarily in order to establish the independence of contractions with it and derive a system 
of linear equations. The tadpole coefficient may been obtained as a solution of this system. 

The general application of our procedure requires further investigation.  Since single cuts of spurious terms do not vanish, 
we need to understand the single cuts of general integrands in order to isolate tadpole coefficients.
We have made a start in this direction in Section~\ref{Sec:SingleGeneralIntegrands}.   It would be very interesting to generalize this analysis.  The simple 
expressions of the terms proportional to $\Lambda^2 t$ lead us to speculate that further results will be similarly simple.  Perhaps there is even a more direct 
way to derive these terms in particular.  

Although
in the examples in Section~\ref{ComTadCoeff}, the system of equations derived from single cuts was sufficient to uniquely determine  the tadpole coefficient,  we need more general information about the system to be sure that it will always work.
The  rank of the matrix associated to the system of linear equation could be insufficient to find a unique solution for the tadpole coefficient.  
A possible concern is that $q$ is null and thus has fewer degrees of freedom than a generic vector.  Here again, study of single 
cuts of general integrands could illuminate the role played by $q$ and perhaps resolve our concern about the general solvability of the linear system.

In Passarino-Veltman reduction, a single system of equations is solved for all coefficients of master integrals, and no spurious terms are involved.  It is also unnecessary to reduce all the way to the tadpole level.  In our procedure, by contrast, we actually solve two systems of equations: the first one is the expansion of the tensor numerator in a suitable basis of independent tensors.  However, the second system leading to the tadpole coefficient is then quite simple and usually reducible.  We have assumed that the coefficients of other master integrals would be obtained by (generalized) unitarity methods, so the total calculation is in fact quite long.  We cannot claim that our method for getting tadpole coefficients will be efficient.  Nevertheless, we have found it illuminating to investigate the single cut integral formalism in general.  It would be very interesting to further probe the analytic structure of amplitudes, for example by developing a $D$-dimensional extension for studying their rational parts along the lines of \cite{NigelGlover:2008ur}.

\section*{Acknowledgments}
We would like to thank P. Mastrolia and T. Robens for discussions inspiring this investigation.  R.B. is supported by the {\it Agence Nationale de la Recherche} under grant ANR-09-CEXC-009-01. E.M. is supported by the European Research Council under Advanced Investigator Grant ERC-AdG-228301.

%\newpage

\appendix

\section{Phase space parametrization}
In this appendix we describe some details of the evaluation  of the single cut.   
The starting point is the integral 
 \be
 \Label{Eq:PMstar}
   \int_{\mathbb{R}^4} d^4 \ell_1   \delta^+(\ell_1^2) \; g,
 \ee
 where $g$ is  a general integrand.
 Given an arbitrary four-vector $K$, such that $K^2 \ne 0$ and its energy component is positive, $K_0>0$, we construct a pair of light-like 
 momenta $p$ and $q$ such that $p+q =K$, and two more momenta $ \epsilon_1$,  $\epsilon_2$ defined as   
 \bd
 \epsilon_1 = \frac{1}{2} \Big ( \langle q |\gamma^{\mu} | p  ]+\langle p |\gamma^\mu | q ]\Big ), ~~~
 \epsilon_2= \frac{1}{2i} \Big ( \langle q |\gamma^{\mu} | p  ]- \langle p |\gamma^\mu | q ] \Big ).
 \ed
 The loop momentum $\ell_1$ can be decomposed in the basis $\{ p,q,\epsilon_1,\epsilon_2\}$, 
 \bd
  \ell_1 = t(p + \alpha q + \epsilon_{1} x -\epsilon_{2}  y ), %\equiv t \ell,
 \ed
 and the integral~(\ref{Eq:PMstar}) can be expressed 
 in terms of the coordinates $(t, \alpha, x, y)$.
 The Jacobian of this reparametrization reads as follows:
 \bd
 \mathcal{J} =  \left |  - t^3 \varepsilon_{\mu \nu \rho \sigma} p^\mu  q^\nu \epsilon_{1}^{\rho} \epsilon_{2}^{ \sigma} \right | = \frac{2 (K^2)^2}{4}  \left | t^3 \right |, 
 \ed
 where in the last step the relation
 \bd
  4 i\; \varepsilon^{\mu \nu \rho \sigma} a_\mu b_{\nu} c_{\rho} d_{\sigma}
  = - 2 \langle a |b\,c\,d| a ] + (2a\cdot b)(2c\cdot d) - (2a\cdot c)(2b \cdot d) +  (2a\cdot d)(2b \cdot c),
 \ed
 has been used~\cite{Dixon:1996wi}. The integral~(\ref{Eq:PMstar})  becomes
 \begin{eqnarray}
   \int d^4 \ell_1  \delta^+(\ell_1^2)\; g &=&
\int  dt \,d x\,d  y\,d \alpha \,  \mathcal{J}
 \delta^+\left (t^2\,  (\alpha -x^2 -y^2 ) \right )  \; g\nonumber \\
 &=&  \int _0^\infty dt ~\int_{\mathbb{R}^2}  d x~d  y~\frac{K^2\,t}{2}\; g.
 \Label{Eq:IntD4L}
 \end{eqnarray}
 This integral is computed using the generalized Cauchy formula, 
  \be
  \label{Eq:CPF}
  2 \pi i \sum_{\mbox{\tiny poles}~z_j} \mbox{Res}\{ F(z,\bar z), z_j\}=
  \oint_{\partial S} F(z,\bar z) dz  - \int_S \frac{\partial F}{\partial \bar z} d\bar z  \wedge d z.
  \ee
  Considering $g$ as a function  of  $(z, \bar z) = (x + i y,  x -i y)$, 
 we construct  its primitive $G(z, \bar z)$  with respect to 
 $\bar z$. Choosing $S$ to be the complex plane $\mathbb{C} $,  we get 
 \be
 \int_{\mathbb{R}^2}  d x\;d  y \; g = \frac{1}{2i}  \int_\mathbb{C} \frac{\partial G}{\partial \bar z} d\bar z  \wedge d z =\frac{1}{2i} \int_{\partial\mathbb{C}} G(z,\bar z) dz - \pi\sum_{\mbox{\tiny poles}~z_j} \mbox{Res}\{ G(z,\bar z), z_j\}. 
 \Label{Eq:IntMain}
 \ee
 In practice we regularize the divergences of the integral~(\ref{Eq:IntMain}) taking 
 instead of  $\mathbb{C} $
 a disk $D$ in the  complex plane enclosing all poles of $G(z, \bar z)$
 viewed as a function of $z$.  In polar coordinates the disk is parametrized as follows,
\bean
z=re^{i\alpha}; \qquad D=\{(r,\alpha) ~|~ 0\leq r \leq \Lambda, ~0 \leq \alpha < 2 \pi \}.
\eean
Under these assumptions, equation (\ref{Eq:IntMain}) becomes
 \be
 \int_{\mathbb{R}^2}  d x\;d  y \; g  = \lim_{\Lambda \to \infty}
 \frac{1}{2} \int_{0}^{2 \pi} \Lambda e^{i \alpha}G\left (\Lambda e^{i \alpha},\Lambda e^{-i \alpha} \right) d\alpha - \pi\sum_{\mbox{\tiny poles}~z_j} \mbox{Res}\{ G(z,\bar z), z_j\}. 
 \Label{Eq:IntMain1}
 \ee
In equations~(\ref{Eq:IntD4L})  and~(\ref{Eq:IntMain})  the correct prefactors are kept, even though in unitarity methods these constant factors cancel out and can be neglected.\footnote{For instance, in the definition of the double cut of Ref.~\cite{Mastrolia:2009dr} and related papers, a factor of  $i/4$ has 
been dropped.  We thank T. Robens for pointing it out.}

\bibliographystyle{JHEP}
\bibliography{references}

\end{document}